\documentclass[12pt]{article}
\usepackage{amsmath}
\usepackage{amssymb}
\usepackage{amsfonts}
\usepackage{graphics}

\newcommand{\field}[1]{\mathbb{#1}}
\newcommand{\N}{\field{N}}
\newcommand{\R}{\field{R}}

\title{Quasi-exactly solvable potentials in Wigner-Dunkl quantum mechanics}

\author{C.\ QUESNE\thanks{E-mail address: christiane.quesne@ulb.be} \\
{\small\sl Physique Nucl\'eaire Th\'eorique et Physique Math\'ematique, 
Universit\'e Libre de Bruxelles,} \\ 
{\small \sl Campus de la Plaine CP229, Boulevard~du Triomphe, B-1050
Brussels, Belgium}}
\date{ }
\begin{document}
\baselineskip=22pt plus 1pt minus 1pt
\maketitle

\begin{abstract}
It is shown that the Dunkl harmonic oscillator on the line can be generalized to a quasi-exactly solvable one, which is an anharmonic oscillator with $n+1$ known eigenstates for any $n\in \N$. It is also proved that the Hamiltonian of the latter can also be rewritten in a simpler way in terms of an extended Dunkl derivative. Furthermore, the Dunkl isotropic oscillator and Dunkl Coulomb potentials in the plane are generalized to quasi-exactly solvable ones. In the former case, potentials with $n+1$ known eigenstates are obtained, whereas, in the latter, sets of $n+1$ potentials associated with a given energy are derived.
\end{abstract}

\vspace{0.5cm}

 
\newpage
%
%
\section{Introduction}

In 1950, Wigner considered a deformed version of the Heisenberg algebra in quantum mechanics \cite{wigner} and then Yang introduced the reflection operator to discuss the harmonic oscillator problem along these lines \cite{yang}. Sets of differential-difference operators associated with finite reflection groups were independently introduced by Dunkl \cite{dunkl89}. These operators, now referred to as Dunkl operators, have proved very useful for studying polynomials in several variables with discrete symmetry \cite{dunkl14}. In physics too, such operators have been employed for bosonizing supersymmetric quantum mechanics \cite{plyu} or generalizations thereof \cite{cq21} and have provided a tool for proving the superintegrability of some models \cite{cq10}.\par
%
%
More recently, deformations of the momentum operator of such a type in quantum mechanics have attracted much attention and have given rise to what is often called Wigner-Dunkl quantum mechanics. For instance, the Dunkl oscillator and the Dunkl-Coulomb problems in the plane were investigated \cite{genest13, genest15}, as well as their generalizations to three dimensions \cite{genest14, gha19}. Coherent states \cite{gha22}, a generalization of shape invariance in supersymmetric quantum mechanics \cite{dong22}, and relativistic effects \cite{mota21, mota19} have also been considered. The possibility of using Dunkl derivatives with two or three parameters has been investigated \cite{dong21, mota22}. Very recently, some rationally-extended potentials, whose wavefunctions can be expressed in terms of exceptional orthogonal polynomials have been studied in the Wigner-Dunkl formalism \cite{cq23a, cq23b}.\par
%
%
All problems considered so far in such a framework being exactly solvable (ES) ones, the purpose of the present work is to extend them to quasi-exactly solvable (QES) problems. In standard quantum mechanics, the latter occupy an intermediate place between ES and non-solvable problems and are characterized by the fact that several eigenstates are explicitly known in connection with a hidden ${\rm sl}(2, \R)$ algebraic structure, while the remaining ones are unknown.\par
%
%
This paper is organized as follows. In the next section, we present a first example of QES potential, which is a particular kind of anharmonic oscillator on the line. In the third and fourth sections, we give some QES potentials that extend the known isotropic and Coulomb potentials in the plane. Finally, in the last section, we summarize our findings.\par
%
%
\section{QES oscillator potential on the line}

The standard Dunkl oscillator model on the line is defined by the Hamiltonian \cite{genest13}
\begin{equation}
  {\cal H}_{\mu} = \frac{1}{2} (- D_{\mu}^2 + x^2) = \frac{1}{2}\left( - \frac{d^2}{dx^2}- \frac{2\mu}{x}
  \frac{d}{dx} + \frac{\mu}{x^2}(1-R) + x^2\right),
\end{equation}
where $D_{\mu}$ is the Dunkl derivative
\begin{equation}
  D_{\mu} = \frac{d}{dx} + \frac{\mu}{x}(1-R), \qquad \mu>-\frac{1}{2},
\end{equation}
and $R$ is the reflection operator, defined by $R f(x) = f(-x)$ and whose eigenvalues are $1-2\epsilon$ with $\epsilon=0$ and $\epsilon=1$ for even and odd functions, respectively. The corresponding ES Schr\"odinger equation 
\begin{equation}
  {\cal H}_{\mu} \psi^{(\mu)}_{2k+\epsilon}(x) = \left(2k+\epsilon+\mu+\frac{1}{2}\right) \psi^{(\mu)}_
  {2k+\epsilon}(x) , \qquad k=0,1,2,\ldots, \qquad \epsilon=0,1,  \label{eq:DHO}
\end{equation}
has wavefunctions $\psi^{(\mu)}_{2k+\epsilon}(x) = \exp(-\frac{1}{2}x^2) H^{(\mu)}_{2k+\epsilon}(x)$, expressed in terms of generalized Hermite polynomials $H^{(\mu)}_{2k+\epsilon}(x)$, which are even (resp.\ odd) for $\epsilon=0$ (resp.\ $\epsilon=1$) and may be defined in terms of Laguerre polynomials as $H^{(\mu)}_{2k+\epsilon}(x) \propto x^{\epsilon} L^{(\mu-1/2+\epsilon)}_k(x^2)$.\par
%
%
These results for the Dunkl oscillator on the line can be directly obtained \cite{cq23a} from well-known ones for the (three-dimensional) radial oscillator, whose Hamiltonian is given by
\begin{equation}
  H_l = - \frac{d^2}{dr^2} + r^2 + \frac{l(l+1)}{r^2},  \label{eq:HO}
\end{equation}
with $l$ denoting the angular momentum quantum number and $r$ running on the half-line $0<r<+\infty$. The corresponding Schr\"odinger equation reads
\begin{equation}
  H_l \phi^{(l)}_k(r) = 2\left(2k+l+\frac{3}{2}\right) \phi^{(l)}_k(r), \quad \phi^{(l)}_k(r) \propto 
  e^{-\frac{1}{2}r^2} r^{l+1} L^{(l+1/2)}_k(r^2), \quad k=0,1,2,\ldots. \label{eq:HO-SE} 
\end{equation}
\par
%
%
On starting from the latter, performing the replacements $r\to x$, $l\to \mu-1$ (resp.\ $l\to \mu$), and multiplying the wavefunction by $x^{-\mu}$, we indeed arrive at (\ref{eq:DHO}) for $\epsilon=0$ (resp.\ $\epsilon=1$) with $\psi^{(\mu)}_{2k}(x) \propto x^{-\mu} \phi^{(\mu-1)}_k(x)$ (resp.\ $\psi^{(\mu)}_{2k+1}(x) \propto x^{-\mu} \phi^{(\mu)}_k(x)$).\footnote{For details about the wavefunctions normalization, including the change of normalization from the half-line $0<r<+\infty$ to the whole line $-\infty < x < +\infty$, see \cite{cq23a}.}\par
%
%
{}From \cite{turbiner88, turbiner16}, it is known that the ES radial oscillator Hamiltonian (\ref{eq:HO}) can be extended to a QES one
\begin{equation}
  H^{(n)}_{a,b,l} = - \frac{d^2}{dr^2} + a^2 r^6 + 2ab r^4 + [b^2 - (2l+5+4n)a] r^2 + \frac{l(l+1)}{r^2},
  \label{eq:HO-ext}
\end{equation}
where $a$, $b$ are real parameters (such that $a>0$) and $n$ is a non-negative integer labelling the corresponding ${\rm sl}(2, \R)$ irreducible representation. For a given $n$ value, the Schr\"odinger equation corresponding to (\ref{eq:HO-ext}),
\begin{equation}
  H^{(n)}_{a,b,l} \phi^{(n)}_{a,b,l,k}(r) = E^{(n)}_{a,b,l,k} \phi^{(n)}_{a,b,l,k}(r),
\end{equation}
had $n+1$ known eigenfunctions
\begin{equation}
  \phi^{(n)}_{a,b,l,k}(r) \propto e^{-\frac{a}{4}r^4 - \frac{b}{2}r^2} r^{l+1} p^{(n)}_k(r^2), \qquad k=0, 1,
  \ldots, n,
\end{equation}
where $p^{(n)}_k(r^2)$ are polynomials of degree $n$ in $r^2$, obtained by diagonalizing an $(n+1)\times(n+1)$ matrix, whose eigenvalues are $E^{(n)}_{a,b,l,k}$.\par
%
%
On applying to (\ref{eq:HO-ext}) the same kind of transformation as above for its ES counterpart, we arrive, for $a>0$, $\mu>-\frac{1}{2}$, and $n\in\N$, at a QES Dunkl oscillator, whose Hamiltonian is given by
\begin{equation}
  {\cal H}^{(n)}_{a,b,\mu} = \frac{1}{2}\{- D_{\mu}^2 + a^2 x^6 + 2ab x^4 + [b^2 - (2\mu + 4 + 4n -R)a] 
  x^2\}
\end{equation}
or
\begin{equation}
  {\cal H}^{(n)}_{a,b,\mu} = \frac{1}{2}\left\{- \hat{D}_{\mu,a}^2 + \frac{8}{9}a^2 x^6 + 2ab x^4 + 
  \left[b^2 - \left(\frac{8}{3}\mu + 4 + 4n\right)a\right] x^2 \right\}
\end{equation}
in terms of an extended Dunkl derivative $\hat{D}_{\mu,a} = D_{\mu} - \frac{1}{3}ax^3 R$. For a given $n$ and every $\epsilon$ value, the corresponding Schr\"odinger equation has $n+1$ known eigenvalues and eigenfunctions, whose explicit expressions can be directly obtained from results given in \cite{turbiner16}. For $n=0$, for instance, we get a known lowest-energy state, such that
\begin{equation}
  {\cal E}^{(a,b,\mu)}_{\epsilon} = \mu + \epsilon + \frac{1}{2}, \qquad \psi^{(a,b,\mu)}_{\epsilon} (x) \propto 
  e^{-\frac{1}{4}ax^4 - \frac{1}{2}bx^2} x^{\epsilon},
\end{equation}
while for $n=1$, we obtain the two lowest-energy states, for which
\begin{align}
  & {\cal E}^{(a,b,\mu)}_{\epsilon,\begin{smallmatrix}0\\1\end{smallmatrix}} = \left(\mu+\epsilon+\frac{3}{2}
       \right)b \mp \sqrt{b^2 + 2a(2\mu+2\epsilon+1)}, \\
  & \psi^{(a,b,\mu)}_{\epsilon, \begin{smallmatrix}0\\1\end{smallmatrix}}(x) \propto e^{-\frac{1}{4}ax^4
       -\frac{1}{2}bx^2} x^{\epsilon} \left(2ax^2 + b \pm \sqrt{b^2 + 2a(2\mu+2\epsilon+1)}\right).
\end{align}
\par
%
%
\section{QES oscillator potential in the plane}

In the plane, one has two Dunkl derivatives $D_{\mu_i} = \frac{d}{dx_i} + \frac{\mu_i}{x_i}(1-R_i)$, $i=1,2$, where $\mu_i>-\frac{1}{2}$ and $R_i$, defined by $R_i f(x_i) = f(-x_i)$, has eigenvalues $1-2\epsilon_i$, with $\epsilon_i=0$ (resp.\ $\epsilon_i=1$) for even (resp.\ odd) functions $f(x_i)$. The corresponding Hamiltonian is therefore ${\cal H}_{\mu_1,\mu_2} = \frac{1}{2}(D_{\mu_1}^2 + D_{\mu_2}^2) + V(x_1,x_2)$.\par
%
%
{}For the case of the isotropic oscillator $V(x_1,x_2) = \frac{1}{2}(x_1^2 + x_2^2)$, the Schr\"odinger equation can be rewritten in polar coordinates $\rho$, $\phi$ (with $0<\rho<+\infty$, $0<\phi<2\pi$), and leads to the radial equation \cite{genest13, cq23b}
\begin{equation}
  \frac{1}{2} \left(- \frac{d^2}{d\rho^2} - \frac{2\mu_1+2\mu_2+1}{\rho} \frac{d}{d\rho} + \rho^2 +
   \frac{M^2}{\rho^2}\right) R(\rho) = {\cal E} R(\rho),  \label{eq:radial}
\end{equation}
with $M^2 = 4\nu(\nu+\mu_1+\mu_2)$, where $\nu=0, 1, 2,\ldots$ if $\epsilon_1=\epsilon_2 =0$, $\nu=1, 2, 3, \ldots$ if $\epsilon_1=\epsilon_2=1$, and $\nu=\frac{1}{2}, \frac{3}{2}, \frac{5}{2}, \ldots$ if $\epsilon_1 =0$, $\epsilon_2=1$ or $\epsilon_1=1$, $\epsilon_2=0$. The change of function $R(\rho) = \rho^{-\mu_1-\mu_2-\frac{1}{2}} Q(\rho)$ transforms (\ref{eq:radial}) into
\begin{equation}
  \frac{1}{2}\left(- \frac{d^2}{d\rho^2} + \rho^2 + \frac{(2\nu+\mu_1+\mu_2-\frac{1}{2})(2\nu+\mu_1
  +\mu_2+\frac{1}{2})}{\rho^2}\right) Q(\rho) = {\cal E} Q(\rho),  \label{eq:radial-bis}
\end{equation}
which is similar to the Schr\"odinger equation for the radial oscillator (\ref{eq:HO}), the role of $l$ being played by $2\nu+\mu_1+\mu_2-\frac{1}{2}$. From this, it directly follows that $\cal E$ and $R(\rho)$ are given by ${\cal E}_{k,\nu} = 2k+2\nu+\mu_1+\mu_2+1$ and $R_{k,\nu}(\rho) \propto e^{-\frac{1}{2}\rho^2} \rho^{2\nu} L^{(2\nu+\mu_1+\mu_2)}_k(\rho^2)$ with $k=0, 1, 2, \ldots$. This radial function is multiplied by an angular function $\Phi^{(\epsilon_1,\epsilon_2)}_{\nu}(\phi)$, whose explicit form is given in \cite{genest13, cq23b}.\par
%
%
The similarity between (\ref{eq:radial-bis}) and the radial oscillator Schr\"odinger equation (\ref{eq:HO-SE}) provides a hint for extending the ES Dunkl oscillator in the plane to a QES one. From (\ref{eq:HO-ext}) with $r\to \rho$, $l\to 2\nu+\mu_1+\mu_2-\frac{1}{2}$, one gets an extension of (\ref{eq:radial-bis}), which after the change of function from $Q(\rho)$ to $R(\rho)$ reads
\begin{align}
  & \frac{1}{2}\biggl\{- \frac{d^2}{d\rho^2} - \frac{2\mu_1+2\mu_2+1}{\rho} \frac{d}{d\rho} + a^2\rho^6
        + 2ab\rho^4 + [b^2 - (4\nu+2\mu_1+2\mu_2+4+4n)a]\rho^2  \nonumber \\
  & \quad {}+ \frac{M^2}{\rho^2}\biggr\} {\cal R}^{(a,b,n)}_{k,\nu}(\rho) = {\cal E}^{(a,b,n)}_{k,\nu}
        {\cal R}^{(a,b,n)}_{k,\nu}(\rho),
\end{align}
where $a>0$, $n\in\N$, and with $n+1$ known solutions for $k=0, 1, \ldots, n$.\par
%
%
{}For $n=0$, for instance, we get a lowest-energy state with
\begin{equation}
  {\cal E}^{(a,b,0)}_{0,\nu} = (2\nu+\mu_1+\mu_2+1) b, \qquad {\cal R}^{(a,b,0)}_{0,\nu}(\rho) \propto
  e^{-\frac{1}{4}a\rho^4 - \frac{1}{2}b\rho^2} \rho^{2\nu},
\end{equation}
while, for $n=1$, we obtain the two lowest-energy states for which
\begin{equation}
\begin{split}
  {\cal E}^{(a,b,1)}_{\begin{smallmatrix}0\\1\end{smallmatrix},\nu} &= (2\nu+\mu_1+\mu_2+2)b \mp 
      \sqrt{b^2 + 2a(4\nu+2\mu_1+2\mu_2+2)}, \\
  {\cal R}^{(a,b,1)}_{\begin{smallmatrix}0\\1\end{smallmatrix},\nu}(\rho) &\propto 
       e^{-\frac{1}{4}a\rho^4-\frac{1}{2}b\rho^2}
      \rho^{2\nu}\left(2a\rho^2+b\pm\sqrt{b^2 + 2a(4\nu+2\mu_1+2\mu_2+2)}\right).
\end{split}
\end{equation}
Note that the angular equation remaining unchanged, these radial wavefunctions are multiplied by an angular wavefunction $\Phi^{(\epsilon_1, \epsilon_2)}_{\nu}(\phi)$, as given in \cite{genest13,cq23b}.\par
%
%
\section{QES Coulomb potential in the plane}

In polar coordinates, the Dunkl-Coulomb potential in the plane \cite{genest15} can be treated in a way rather similar to that of the oscillator, the potential $\frac{1}{2}\rho^2$ being replace by $- \frac{\alpha}{2\rho}$ (with $\alpha>0$) in (\ref{eq:radial}) and (\ref{eq:radial-bis}).\footnote{The potential $\alpha/\rho$ of \cite{genest15} is replaced here by $-\alpha/(2\rho)$ to conform with the choice made in \cite{turbiner16}.} The counterpart of (\ref{eq:radial-bis}) looks like the radial equation of the (three-dimensional) Coulomb problem with Hamiltonian
\begin{equation}
  H_l = - \frac{d^2}{dr^2} - \frac{\alpha}{r} + \frac{l(l+1)}{r^2},  \label{eq:Coulomb}
\end{equation}
$l$ denoting the angular momentum quantum number and $r$ running on the half-line $0<r<+\infty$. With $r$ replaced by $\rho$ and $l$ by $2\nu+\mu_1+\mu_2-\frac{1}{2}$, one gets for the Dunkl Coulomb problem in the plane ${\cal E}_{k,\nu} = -\alpha^2/[8(k+2\nu+\mu_1+\mu_2+1/2)]^2$, $R_{k,\nu}(\rho) \propto e^{-\frac{1}{2}\beta\rho} (\beta\rho)^{2\nu} L^{(4\nu+2\mu_1+2\mu_2)}_k(\beta\rho)$ with $\beta=\sqrt{-8{\cal E}_{k,\nu}}$, $k=0,1,2,\ldots$, and the same angular wavefunction $\Phi^{(\epsilon_1,\epsilon_2)}_{\nu}(\phi)$ as before.\par
%
%
{}From \cite{turbiner88, turbiner16}, it is known that for any real parameters $a$, $b$ with $a>0$ and any $n\in\N$, the ES Coulomb Hamiltonian (\ref{eq:Coulomb}) can be extended to a set of $n+1$ QES Hamiltonians, given by
\begin{equation}
  H^{(n)}_{a,b,l,k} = -\frac{d^2}{dr^2} + a^2 r^2 + 2abr - \frac{b(2l+2)+\alpha_k}{r} + \frac{l(l+1)}{r^2},
  \qquad k=0,1,\ldots, n,  \label{eq:Coulomb-ext}
\end{equation}
where $\alpha_0 > \alpha_1 > \cdots > \alpha_n$. The known eigenstates
\begin{equation}
  \phi^{(n)}_{a,b,l,k}(r) \propto e^{-\frac{1}{2}ar^2-br} r^{l+1} p^{(n)}_k(r), \qquad k=0,1,\ldots,n,
  \label{eq:p}
\end{equation}
are characterized by a fixed energy $E^{(n)}_{a,b,l} = a(2n+2l+3) - b^2$ and are the ground state for the Hamiltonian with $k=0$, the first-excited state for that with $k=1$, \ldots, the $n$th-excited state for that with $k=n$. In (\ref{eq:p}), $p^{(n)}_k(r)$, $k=0,1,\ldots,n$, are $n$th-degree polynomials in $r$, resulting from the diagonalization of an $(n+1)\times(n+1)$ matrix, whose eigenvalues are $\alpha_k$, $k=0,1,\ldots,n$.\par
%
%
By proceeding as in the oscillator case and changing $r$ into $\rho$ and $l$ into $2\nu+\mu_1+\mu_2-\frac{1}{2}$, one can get from (\ref{eq:Coulomb-ext}) a set of radial equations for $n+1$ QES Coulomb potentials
\begin{align}
  & \frac{1}{2}\biggl\{- \frac{d^2}{d\rho^2} - \frac{2\mu_1+2\mu_2+1}{\rho} \frac{d}{d\rho} + a^2\rho^2
        + 2ab\rho - \frac{b(4\nu+2\mu_1+2\mu_2+1)+\alpha_k}{\rho}  \nonumber \\
  & \quad {}+ \frac{M^2}{\rho^2}\biggr\} {\cal R}^{(a,b,n)}_{k,\nu}(\rho) = {\cal E}^{(a,b,n)}_{\nu}
        {\cal R}^{(a,b,n)}_{k,\nu}(\rho), \qquad k=0,1,\ldots,n,
\end{align}
with ${\cal E}^{(a,b,n)}_{\nu} = a(n+2\nu+\mu_1+\mu_2+1) - \frac{1}{2}b^2$.\par
%
%
{}For $n=0$, for instance, we get a single extended potential corresponding to $\alpha_0=0$ and its known lowest-energy state with
\begin{equation}
  {\cal E}^{(a,b,0)}_{\nu} = a(2\nu+\mu_1+\mu_2+1) - \frac{1}{2}b^2, \qquad {\cal R}^{(a,b,0)}_{0,\nu}
  (\rho) \propto e^{-\frac{1}{2}a\rho^2-b\rho} \rho^{2\nu}.
\end{equation}
For $n=1$, we obtain two different potentials corresponding to
\begin{equation}
  \alpha_{\begin{smallmatrix}0\\1\end{smallmatrix}} = b \pm \sqrt{b^2 + 2a(4\nu+2\mu_1+2\mu_2+1)},
\end{equation}
and for that associated with $\alpha_0$ (resp.\ $\alpha_1$) a lowest-energy (resp.\ first-excited) state, whose wavefunctioin is given by
\begin{equation}
  {\cal R}^{(a,b,1)}_{\begin{smallmatrix}0\\1\end{smallmatrix}, \nu}(\rho) \propto e^{-\frac{1}{2}a\rho^2
  -b\rho} \rho^{2\nu}\left(2a\rho+b \pm \sqrt{b^2+2a(4\nu+2\mu_1+2\mu_2+1)}\right), 
\end{equation}
with the same energy
\begin{equation}
  {\cal E}^{(a,b,1)}_{\nu} = a(2\nu+\mu_1+\mu_2+2) - \frac{1}{2}b^2
\end{equation}
in both cases. Note again that the angular equation remaining unchanged, these radial wavefunctions are multiplied by an angular wavefunction $\Phi^{(\epsilon_1,\epsilon_2)}_{\nu}(\phi)$.\par
%
%
\section{Conclusion}

In this work, by using some known results in standard quantum mechanics, we have shown that it is possible to extend some ES potentials into QES ones in Wigner-Dunkl quantum mechanics too.\par
%
%
{}First, on using its relation to the standard radial oscillator, the Dunkl harmonic oscillator on the line has been generalized to a QES anharmonic one, which has been rewritten in a simpler way in terms of an extended Dunkl derivative.\par
%
%
Second, the Dunkl isotropic oscillator and the Dunkl-Coulomb potentials in the plane have been generalized by using the similarity between their radial equation and that of the standard isotropic oscillator or standard Coulomb potential, respectively. As a result, we have obtained for $n=0,1,2,\ldots$, QES Dunkl oscillator-like potentials with $n+1$ known eigenstates, as well as sets of $n+1$ QES Dunkl Coulomb-like potentials with a given known energy.\par
%
%
\section*{Acknowledgment}

The author was supported by the Fonds de la Recherche Scientifique-FNRS under Grant No.\ 4.45.10.08.\par
%
%
\section*{Data availability statement}

No new data were created or analyzed in this study.\par
%
%

 \end{document}